\documentclass{IEEEtran}
\IEEEoverridecommandlockouts
\usepackage{cite}
\usepackage{amsmath,amssymb,amsfonts}
\usepackage{graphicx}
\usepackage{textcomp}
\usepackage{xcolor}
\usepackage{amsthm}
\usepackage{geometry}
\usepackage{float}
\usepackage{algorithm}
\usepackage[noend]{algorithmic}
\usepackage{graphicx}
\usepackage{adjustbox}
\def\BibTeX{{\rm B\kern-.05em{\sc i\kern-.025em b}\kern-.08em
    T\kern-.1667em\lower.7ex\hbox{E}\kern-.125emX}}
\begin{document}

\title{GNN: Graph Neural Network and Large Language Model for Data Discovery
}

\author{\IEEEauthorblockN{ Thomas Hoang}\\
\IEEEauthorblockA{\textit{Department of Computer Science} \\
\textit{Denison University}\\
Ohio,  USA\\
hoang$\_$t2@denison.edu}
}
\maketitle

\begin{abstract}
Our algorithm GNN: Graph Neural Network and Large Language Model for Data Discovery inherit the benefits of \cite{hoang2024plod} (PLOD: Predictive Learning Optimal Data Discovery), \cite{Hoang2024BODBO} (BOD: Blindly Optimal Data Discovery) in terms of overcoming the challenges of having to predefine utility function and the human input for attribute ranking, which helps prevent the time-consuming loop process. In addition to these previous works, our algorithm GNN leverages the advantages of graph neural networks and large language models to understand text type values that cannot be understood by PLOD and MOD, thus making the task of predicting outcomes more reliable. GNN could be seen as an extension of PLOD in terms of understanding the text type value and the user's preferences based on not only numerical values but also text values, making the promise of data science and analytics purposes.
\end{abstract}


\section{Introduction}
Understanding valuable insights from large datasets is important in making decisions in this fast-changing world, especially in data science and predictive applications. In detail, data scientists and analysts often aim to identify high-quality data candidates—free of missing values, duplicates, and inconsistencies—before aggregating datasets with diverse attributes for analysis. Taking housing price prediction as an example, in predicting housing prices, various factors come into play, including the location of the property (proximity to urban centers, crime rates), property characteristics (size, style, modernity), and regional policies (tax implications). With their domain-specific knowledge, analysts/scientists rank these attributes to inform the predictive utility function, which machine learning models then use to forecast housing prices.

Consider a scenario where a scientist must prioritize specific attributes over others based on their expertise. For example, an apartment in a central city like New York, with low crime rates, modern amenities, and high safety standards, might be favored over a historic house in a remote area. Even within similar urban settings, choices become complex—such as deciding between a centrally located apartment and a slightly less expensive one in a nearby suburb. Factors like neighborhood friendliness or environmental tranquility can also influence decision-making. Thus, making these variations in attribute importance are captured by a utility function, which quantifies the significance of each attribute in predicting outcomes like housing prices.

Traditional machine learning algorithms often require users to predefine their utility functions, which can be impractical or challenging for many users. The Blindly Optimal Data Discovery (BOD) and Predictive Learning Optimal Data Discovery (PLOD) algorithms \cite{Hoang2024BODBO} \cite{hoang2024plod} address this by asking users to rank attributes and filtering datasets based on these rankings. But these approaches have limits in a way that may still result in subsets of data that do not perfectly align with the user's intended utility function due to the lack of a precise match between the predicted and actual utility functions.

We introduce GNN (Graph Neural Networks and Large Language Models for Data Discovery) to overcome these limitations. GNN enhances the process by utilizing user rankings and employing advanced machine learning techniques, specifically, PLOD prediction for numerical data, Graph Neural Networks (GNNs), and Large Language Models (LLMs) for textual data. These models collaboratively estimate the utility function, improving the prediction accuracy and relevance of the outputted data subsets. As seen as a significant advancement over previous methods by integrating sophisticated models that can better capture and apply the complex relationships inherent in large, mixed-type datasets, GNN offers a promise for a new way of multimodal predicting applications.

\begin{figure}
  \centering
  \includegraphics[width=\linewidth]{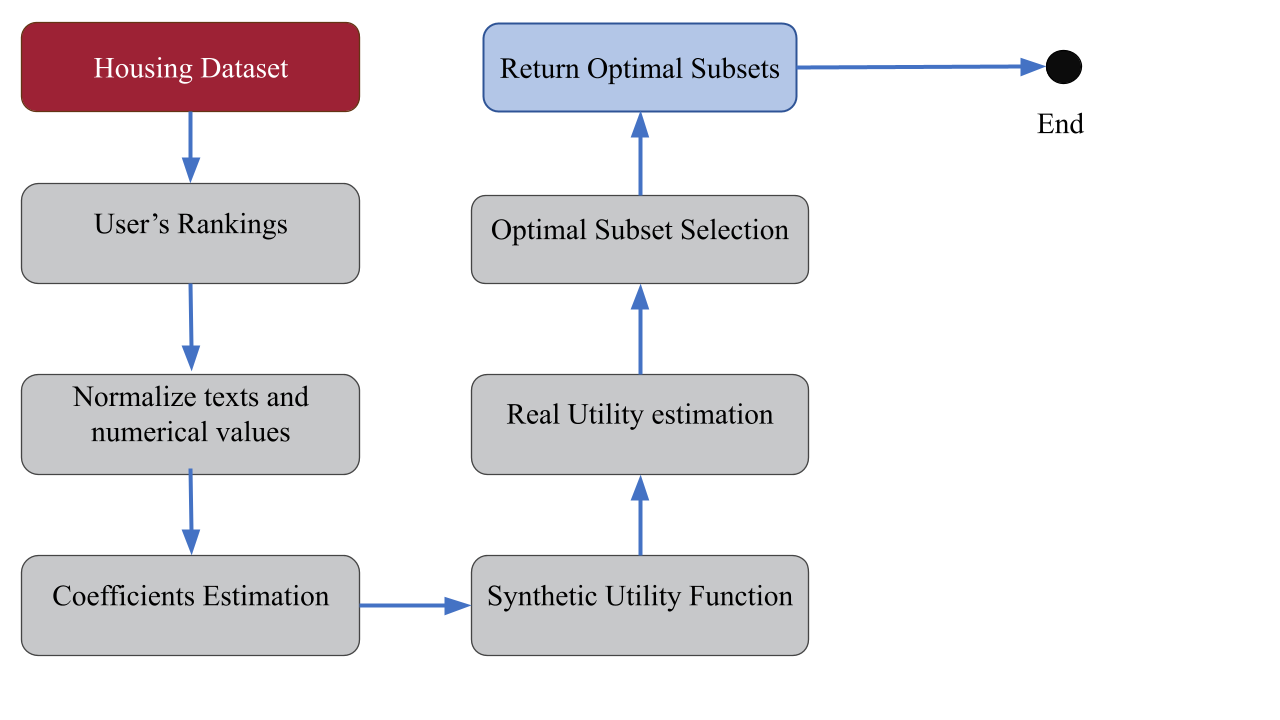}
  \caption{Flowchart of GNN: Graph Neural Networks and Large Language Models for Data Discovery}
  \label{Flowchart}
\end{figure}

\textbf{The work of algorithms}: The workflow as shown in the flowchart \ref{Flowchart}, the algorithm asks the user to rank all the attributes of the dataset then it scales down all the values in the range of $\{0,1\}$ based on the highest to the lowest ranked attributes. Then, GNN will estimate the coefficients of all attributes by leveraging the advantages of Machine Learning for numerical data values and Graph Neural Network and Large Language Model Based on text values to convert all of these values into a score of the range $\{0,1\}$. Then, the process continues with estimating synthetic utility function based on the previous estimation of coefficients. After that, the GNN will estimate the real utility function's coefficients based on the synthetic utility function. Finally, the estimated real utility function 
will be used to output all the subsets of optimal tuples. The workflow as in the Flowchart.

\section{Related Studies}

\textbf{Machine Learning:} The data discovery technique (Metam) \cite{Galhotra2023MetamGD} asks users for user-defined utility functions as input, which has limits that often fall short in scenarios where users may not have a clear understanding or awareness of their utility functions. Other studies using machine learning, including \cite{Abadi2016TensorFlowAS}, \cite{Aken2021AnII}, and \cite{Pavlo2022ExternalVI} like the previous studies of \cite{Galhotra2023MetamGD} about utility functions, which can be limiting in contexts like housing predictions.

\textbf{Top-$k$ Algorithms:} several works such as \cite{FM12}, \cite{LYH09}, \cite{LC09}, \cite{QY12}, and \cite{SI07}, have focused on developing and applying top-$k$ algorithms. Like previous methods, these algorithms require users to specify a utility function, which is not always practical in real-world applications.

\textbf{Skyline Queries:} Several Skyline algorithms have been studied \cite{LYH09}, \cite{LY07}, \cite{MC09}, \cite{CJ06}, \cite{CJ062}, \cite{TX07}, \cite{TD09}, \cite{PT05}, \cite{GY05}, and \cite{XZ08} produce smaller subsets of data from larger datasets but the subsets can still be sizable, potentially including tuples that may not interest the user. To address this, \cite{MLT21} has combined top-$k$ and Skyline techniques to manage the output size better.

\textbf{Multi-objective Optimization:} The work of Deb et al. \cite{deb2002fast} aims to solve problems involving multiple conflicting objectives by identifying solutions that represent various trade-offs (Pareto optimal solutions). Like other optimization techniques, this method presupposes that users can clearly define their goals and utility functions. This assumption often does not hold true, making it less practical for real-world scenarios where user preferences may be implicit or difficult to express.

\textbf{Blindly Optimal Data Discovery:} The Blindly Optimal Data Discovery (BOD) algorithm \cite{Hoang2024BODBO} offers the benefit of its underlying goal without requiring explicit knowledge of the utility function but the limit is that it may still produce subsets that are less relevant for predictive tasks. To better predict outcomes, the Predictive Learning Optimal Data Discovery (PLOD) algorithm incorporates machine learning techniques to estimate the utility function, which achieves its goal of having more accurate and relevant outputs.

\textbf{Graph Neural Networks (GNNs).} Graph Neural Networks have been studied for various tasks as described in the work by Kipf and Welling \cite{kipf2017semi} \cite{Hamilton2017InductiveRL} on semi-supervised learning using GNNs introduced the concept of propagating information through the graph to aggregate node features. However, GNNs' effectiveness relies heavily on the quality of the graph structure and the features used, which can be a limitation when dealing with datasets containing numerical and text features. In addition to this, the work of Velickovic et al. \cite{velickovic2018graph} introduced attention mechanisms in GNNs to improve the weighting of node features, which enhanced GNNs' ability to capture complex dependencies in graph data.

\textbf{Large Language Models (LLMs).} BERT \cite{devlin2019bert} and GPT \cite{brown2020language} have helped to generate human-like text. Still, they have limits that LLMs often Fail to incorporate structured numerical data effectively, which limits their application in domains where both text and numerical data are essential.

\textbf{Combination of GNN and LLM.} By taking advantage of the graphs and language models, we aim to handle both structured data (such as graphs) and unstructured data (like text). Additionally, Wu et al. \cite{wu2021graph} demonstrated the effectiveness of Graph Neural Networks models in tasks like recommendation systems and knowledge graph completion. 

\textbf{Predictive Learning Optimal Data Discovery (PLOD).} the work of \cite{hoang2024plod} takes a predictive approach to optimal data discovery by estimating the user's utility function using machine learning techniques, which could say that, unlike traditional approaches require users to define their utility functions, PLOD leverages machine learning to predict the most relevant tuples based on user preferences. This method has shown promise in applications such as housing market predictions, where user preferences are often implicit and difficult to articulate. However, while PLOD improves upon earlier methods by incorporating predictive modeling, it still relies heavily on the accuracy of the machine learning model, which can influence the quality and quantity of available data.

\textbf{Graph Neural Networks and Large Language Models (GNN).} As we want to take advantage of PLOD and Graph neural Networks'(GNNs) and Large Language Models (LLMs), the GNN algorithm shows the strengths of both GNNs and LLMs, and PLOD to handle datasets, including numerical and text features, by asking users to rank attributes and then leveraging GNNs and LLMs, and PLOD to estimate utility function coefficients. So, GNN offers a novel approach to data discovery that does not require explicit utility function definitions from the user. This method improves upon PLOD by incorporating advanced feature extraction and relational modeling capabilities, making it more robust to the complexities of multimodal data. Thus, GNN offers promising and reliance predicting data science and analytics applications.
\section{Problem Definition}
\begin{table}[htbp]
  \caption{Notation and Meaning for GNN Algorithm}
  \label{tab:GNN_notation}
  \centering
  \begin{adjustbox}{width=\columnwidth}
  \begin{tabular}{|c|l|}
  \hline
  \textbf{Notation} & \textbf{Meaning} \\ 
  \hline
  $T$ & The set of all tuples \\ 
  $D$ & The dataset with both numerical and textual features \\ 
  $X$ & The set of all attributes (numerical and textual) in $D$ \\ 
  $X^{\text{num}}$ & The set of numerical attributes in $D$ \\ 
  $X^{\text{text}}$ & The set of textual attributes in $D$ \\ 
  $x_j^{\text{num}}$ & The $j$th numerical attribute in $X^{\text{num}}$ \\ 
  $x_k^{\text{text}}$ & The $k$th textual attribute in $X^{\text{text}}$ \\ 
  $\beta_{j,\text{num}}$ & Coefficient for the $j$th numerical attribute \\ 
  $\beta_{k,\text{text}}$ & Coefficient for the $k$th textual attribute \\ 
  $h_k^{\text{text}}$ & Feature embedding for the $k$th textual attribute generated by LLM \\ 
  $h_k^{\text{GNN}}$ & Combined feature embeddings processed by GNN \\ 
  $u(x)$ & Utility function estimating the score for a tuple \\ 
  $u_{\text{syn}}(x)$ & Synthetic utility function based on initial coefficient estimates \\ 
  $u_{\text{real}}(x)$ & Final utility function used for optimal tuple selection \\ 
  $t$ & A subset of $T$ containing optimal tuples \\ 
  \hline
  \end{tabular}
  \end{adjustbox}
\end{table}

\textbf{Linear function}: 
\begin{itemize}
    \item $LINEAR = \{f|f(x) =  \sum\limits_{i=1}^{d}f(x_i) \\ \text{\{where each} \ f(x_i)  \text{ is a linear function.}\}$
\end{itemize}

\section*{Mathematical Proof}

We want to take the advantages of PLOD and Graph Neural Networks and Large language Models so the GNN algorithm combines numerical and textual data processing using the advantage of PLOD as in \cite{hoang2024plod}, Graph Neural Networks (GNNs) as in several studies \cite{kipf2017semi} \cite{Hamilton2017InductiveRL} \cite{velickovic2018graph}, and Large Language Models (LLMs) as in \cite{devlin2019bert} \cite{brown2020language}, thus, to estimate the coefficients of a utility function that ranks the importance of different attributes to identify optimal subsets of data.

\subsection{Model Representation}

Below is the utility function:
\[
u(\mathbf{x}) = \sum_{j=1}^{m} \beta_{j,\text{num}} \cdot x_j^{\text{num}} + \sum_{k=1}^{n} \beta_{k,\text{text}} \cdot x_k^{\text{text}} + \epsilon
\]
where:
\begin{itemize}
    \item \( \mathbf{x} = (x_1^{\text{num}}, \ldots, x_m^{\text{num}}, x_1^{\text{text}}, \ldots, x_n^{\text{text}}) \) is the vector of attributes, with \( x_j^{\text{num}} \) representing numerical attributes and \( x_k^{\text{text}} \) representing textual attributes.
    \item \( \beta_{j,\text{num}} \) and \( \beta_{k,\text{text}} \) are the coefficients for numerical and textual attributes, respectively.
    \item \( \epsilon \) is the error term.
\end{itemize}

\subsection{Goal}

We want to estimate the coefficients \( \beta_{j,\text{num}} \) and \( \beta_{k,\text{text}} \) for the utility function \( u(\mathbf{x}) \), combining both numerical and textual data, to predict the utility score.

\subsection{Numerical Data Processing}

For the numerical attributes, we want to apply linear regression techniques to estimate the coefficients \( \beta_{j,\text{num}} \). to minimize the sum of squared errors (SSE) between the observed utility scores and those predicted by the model, the model is trained as below:
\[
SSE_{\text{num}} = \sum_{i=1}^{N} (u_i - \hat{u}_i^{\text{num}})^2 = \sum_{i=1}^{N} \left( u_i - \sum_{j=1}^{m} \beta_{j,\text{num}} \cdot x_{i,j}^{\text{num}} \right)^2
\]
Where \( N \) is the number of data points.

\subsection{Textual Data Processing}

For the textual attributes, we try to estimate the coefficients \( \beta_{k,\text{text}} \). Meanwhile, we let the GNN process the graph structure of the dataset, where each node represents an attribute, and each edge represents a relationship between attributes. LLM processes textual data and outputs feature embeddings that are fed into the GNN:
\[
\mathbf{h}_k^{\text{text}} = \text{GNN}(\text{LLM}(x_k^{\text{text}}))
\]
Our goal is to minimize the SSE for textual attributes:
\[
SSE_{\text{text}} = \sum_{i=1}^{N} (u_i - \hat{u}_i^{\text{text}})^2 = \sum_{i=1}^{N} \left( u_i - \sum_{k=1}^{n} \beta_{k,\text{text}} \cdot h_{i,k}^{\text{text}} \right)^2
\]

\subsection{Synthetic Utility Function}

In order to refine the model, we first estimate a synthetic utility function \( u_{\text{syn}}(\mathbf{x}) \) based on the combined coefficients \( \beta_{j,\text{num}} \) and \( \beta_{k,\text{text}} \):
\[
u_{\text{syn}}(\mathbf{x}) = \sum_{j=1}^{m} \beta_{j,\text{num}} \cdot x_j^{\text{num}} + \sum_{k=1}^{n} \beta_{k,\text{text}} \cdot h_k^{\text{text}}
\]

\subsection{Real Utility Function}

The final utility function \( u_{\text{real}}(\mathbf{x}) \) is then estimated using the synthetic utility function as shown below:
\[
u_{\text{real}}(\mathbf{x}) = \sum_{j=1}^{m} \gamma_{j,\text{num}} \cdot x_j^{\text{num}} + \sum_{k=1}^{n} \gamma_{k,\text{text}} \cdot h_k^{\text{text}} + \delta
\]
where \( \gamma_{j,\text{num}} \) and \( \gamma_{k,\text{text}} \) are the refined coefficients, and \( \delta \) is the error term.

\subsection{Final goal}

Our main goal is to identify the optimal subsets \( t \subset T \) that maximize the real utility function \( u_{\text{real}}(\mathbf{x}) \):
\[
\text{Optimal Subsets} = \arg\max_{t \subset T} \sum_{i \in t} u_{\text{real}}(\mathbf{x}_i)
\]

\begin{algorithm}
\caption{GNN: Graph Neural Networks and Large Language Models for Data Discovery}
\label{GNN Algorithm}
\begin{algorithmic}[1]  
\REQUIRE Dataset $D$ with features $X = \{x_1, x_2, \ldots, x_n\}$ (both numerical and text), and labels $y$
\ENSURE Subsets of optimal tuples $t \subseteq T$ with utility scores

\STATE \textbf{User Ranking:} Ask the user to rank all attributes $x_i \in X$
\FOR{each attribute $x_i \in X$}
    \STATE Scale down values of $x_i$ to the range $\{0,1\}$ based on ranking
\ENDFOR

\STATE \textbf{Coefficient Estimation:}
\STATE Train a Machine Learning model on numerical data to estimate coefficients $\beta_{num}$
\STATE Use Graph Neural Network (GNN) and Large Language Model (LLM) on text data to estimate coefficients $\beta_{text}$
\STATE Combine $\beta_{num}$ and $\beta_{text}$ to form overall coefficients $\beta = \{\beta_{num}, \beta_{text}\}$

\STATE \textbf{Synthetic Utility Function:}
\STATE Estimate synthetic utility function $U_{syn}$ based on coefficients $\beta$

\STATE \textbf{Real Utility Function:}
\STATE Refine and estimate real utility function $U_{real}$ coefficients based on $U_{syn}$

\STATE \textbf{Optimal Subsets Selection:}
\STATE Use $U_{real}$ to score and filter the dataset $D$ to find subsets of optimal tuples $t$

\STATE \textbf{Return} Optimal subsets $t$

\end{algorithmic}
\end{algorithm}

\section*{GNN: Graph Neural Networks and Large Language Models for Data Discovery}

In \textbf{lines 1-3:}, we are gonna let the GNN algorithm ask the user to rank all attributes in the dataset $X$, which contains numerical and textual data. Based on the user's ranking, the attributes are scaled down to a normalized range $\{0, 1\}$. In the \textbf{lines 4-8:}, We then try to estimate the coefficients for each attribute $\beta_{\text{num}}$ for numerical data. After this process, we then try to estimate the coefficients $\beta_{\text{text}}$ for textual data. Following, the algorithm then combines these estimated coefficients into a comprehensive set $\beta = \{\beta_{\text{num}}, \beta_{\text{text}}\}$ that shows the overall importance of each attribute. In the following \textbf{lines 9-10:} After that of using the combined coefficients, the algorithm tries to estimate a synthetic utility function $U_{\text{syn}}$, which is a step designed to approximate the actual utility function. Next in the \textbf{lines 11-12:}, GNN tries to refine this synthetic utility function to produce the actual utility function $U_{\text{real}}$. Next in \textbf{lines 13-14:} Finally, we then try with the real utility function, applied to score and filter the dataset $D$ to identify subsets of optimal tuples $t$. Afterall, in \textbf{line 15:} The algorithm finally returns the optimal subsets $t$ as the final output.

\section{Experimental Analysis}

We will conduct experimental evaluation among various algorithms that return the optimal subset of tuples, which includes the algorithms' precision and runtime performance, including PLOD, GNN, BOD, Top-K, Skyline, and Multi-objective optimization. While conducting experiments, we performed the experiments assuming no errors from the scientist's domain knowledge, meaning that the dataset used consisted of valid integers, duplicates, and missing values. All experiments were executed using Google Colab on a laptop GL 65 Leopard 10SCXK, an x64-based PC, running Microsoft Windows 11 Home Single Language.

\subsection{Runtime Comparison}

\begin{figure}[H]
  \centering
  \includegraphics[width=\linewidth]{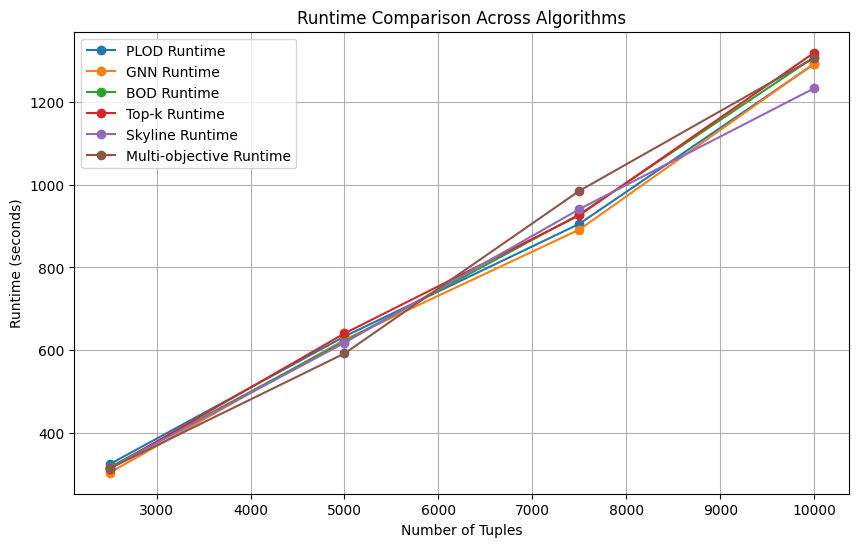}
  \caption{Runtime comparison with changes in number of tuples.}
  \label{Runtime comparison with changes in number of tuples}
\end{figure}

We show in the figure \ref{Runtime comparison with changes in number of tuples} that all algorithms' runtime increases linearly with the number of tuples. Among the algorithms, the GNN exhibited a slightly lower runtime than others. The BOD algorithm demonstrated a relatively higher runtime due to its complex decision-making process. In contrast, the Multi-objective and Skyline algorithms performed comparably, showing similar runtime patterns, while the Top-K algorithm had a runtime close to PLOD.

\subsection{Precision Comparison Across Algorithms}
In the \ref{Precision comparison with changes in number of tuples}, GNN and PLOD exhibited the highest precision, with GNN maintaining a more consistent precision across different tuple sizes. In contrast, BOD, Top-K, and Skyline algorithms demonstrated low precision across all tuple sizes, with little to no improvement as the dataset grew. The Multi-objective algorithm showed moderate precision but decreased as the number of tuples increased.

\begin{figure}[H]
  \centering
  \includegraphics[width=\linewidth]{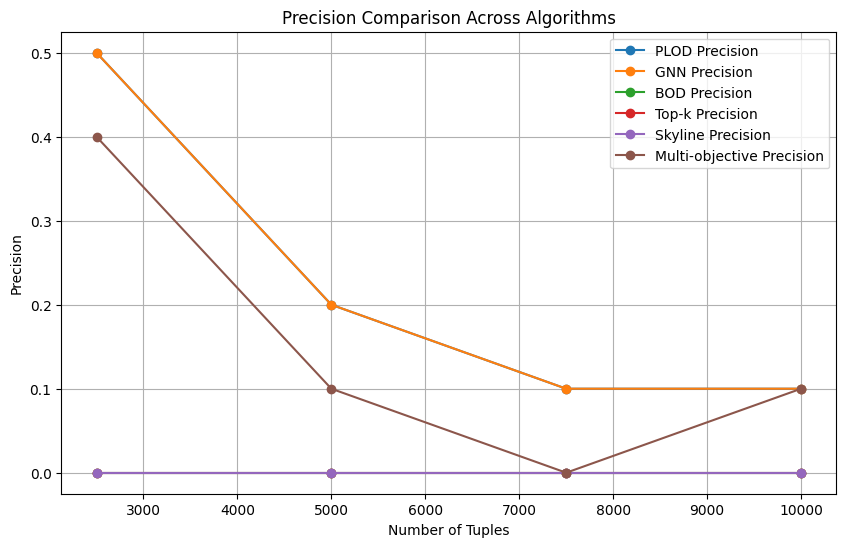}
  \caption{Precision comparison with changes in number of tuples.}
  \label{Precision comparison with changes in number of tuples}
\end{figure}

\subsection{Precision Comparison Between PLOD and GNN Models}

We show in figure \ref{Precision comparison with changes in number of tuples}, the GNN consistently outperformed PLOD in precision across all tuple sizes, which suggests that this technique of combining Graph Neural Networks and Large Language Models provides a more robust predictive capability, particularly as the dataset size increases.

\begin{figure}[H]
  \centering
  \includegraphics[width=\linewidth]{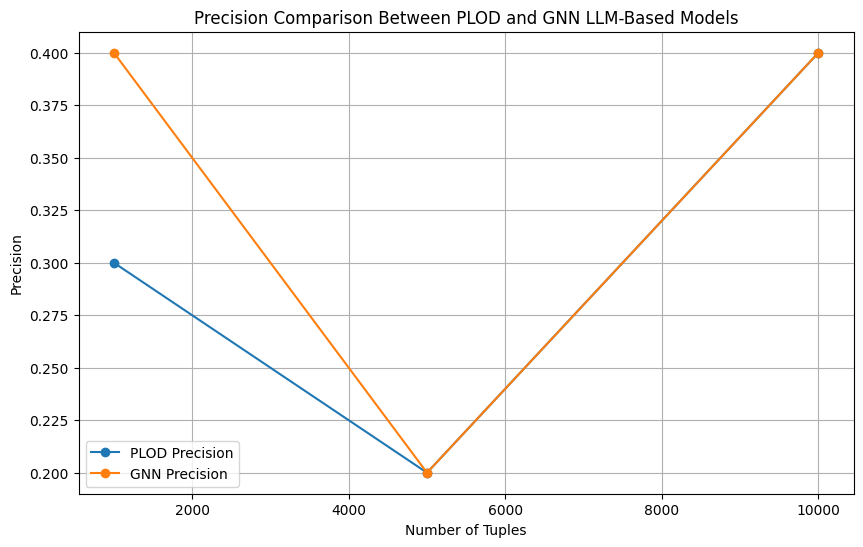}
  \caption{Precision comparison with changes in number of tuples.}
  \label{Precision comparison with changes in number of tuples}
\end{figure}

\section*{Analysis of Precision Scores Across Dataset Sizes}

The PLOD and GNN demonstrate high initial precision scores of 0.5 with smaller datasets, but the performance declines significantly as the dataset size increases, with both models reaching a precision score of 0.1 for larger datasets. In addition, the BOD, Top-K, and Skyline algorithms consistently produce precision scores of 0 across all dataset sizes, indicating their incompetence in identifying relevant tuples within the given datasets. Also, the multi-objective optimization algorithm shows some promise, with an initial precision score of 0.4, but it also declines in performance as the dataset size increases.

\subsection{Precision and Stability Comparison Using Boston Housing Data \cite{nair2021boston}}

We use the Boston Housing dataset as shown in the results summarized in Tables \ref{tab:precision_scores}, \ref{table:stability_precision_analysis}, and \ref{tab:runtime-comparison}.

\begin{table}[H]
\centering
\begin{tabular}{|l|c|}
\hline
\textbf{Algorithm} & \textbf{Precision Score} \\
\hline
PLOD & 0.5 \\
\hline
GNN & 0.6 \\
\hline
BOD & 0.0 \\
\hline
Top-K & 0.1 \\
\hline
Skyline & 0.0 \\
\hline
Multi-objective & 0.4 \\
\hline
\end{tabular}
\caption{Precision Scores on the Boston Housing Dataset}
\label{tab:precision_scores}
\end{table}

Table \ref{tab:precision_scores} shows that the GNN algorithm achieved the highest precision score of 0.6, followed by PLOD with a score of 0.5. The BOD and Skyline algorithms showed no precision, with a score of 0.0, while the Top-K algorithm had a low precision of 0.1. The Multi-objective algorithm had a moderate precision score of 0.4.

\begin{table}[H]
\centering
\begin{tabular}{|l|c|c|}
\hline
\textbf{Algorithm}       & \textbf{Stability} & \textbf{Std Dev} \\ \hline
PLOD                    & 0.570                             & 0.110            \\ \hline
GNN                     & 0.610                             & 0.104            \\ \hline
BOD                     & 0.000                             & 0.000            \\ \hline
Top-K                   & 0.030                             & 0.046            \\ \hline
Skyline                 & 0.040                             & 0.066            \\ \hline
Multi-objective         & 0.490                             & 0.054            \\ \hline
\end{tabular}
\caption{Stability Precision Analysis Results for Different Algorithms}
\label{table:stability_precision_analysis}
\end{table}

Table \ref{table:stability_precision_analysis} indicates that the GNN algorithm is the most stable, with a stability score of 0.610 and a standard deviation of 0.104. PLOD follows closely with a stability score of 0.570 and a standard deviation of 0.110. The BOD algorithm shows no stability, while Top-K and Skyline have low stability scores. The Multi-objective algorithm has a moderate stability score of 0.490.

\begin{table}[H]
\centering
\begin{tabular}{|l|c|}
\hline
\textbf{Algorithm}         & \textbf{Runtime (seconds)} \\ \hline
PLOD                       & 4.7114                     \\ \hline
GNN                      & 5.0887                     \\ \hline
BOD                        & 3.1400                     \\ \hline
Top-K                      & 2.4870                     \\ \hline
Skyline                    & 2.8929                     \\ \hline
Multi-objective            & 3.1718                     \\ \hline
\end{tabular}
\caption{Runtime Comparison Across Algorithms}
\label{tab:runtime-comparison}
\end{table}

We show as in the table \ref{tab:runtime-comparison}, the GNN algorithm, while achieving higher precision and stability, also required the most time to run (5.0887 seconds). PLOD had a slightly lower runtime at 4.7114 seconds. The BOD algorithm had the lowest runtime at 3.1400 seconds, while the Top-K, Skyline, and Multi-objective algorithms had similar runtimes around 3 seconds.

\subsection{Precision and Stability Comparison Using Kaggle Housing Data \cite{housing_price_prediction_kaggle}}

We use the Kaggle Housing dataset in the results summarized in Tables \ref{tab:precision_scores_kaggle}, \ref{tab:stability_analysis_kaggle}, and \ref{table:runtime_comparison_kaggle}.

\begin{table}[H]
\centering
\begin{tabular}{|l|c|}
\hline
\textbf{Algorithm}            & \textbf{Precision Score} \\ \hline
PLOD Precision                & 0.6                      \\ \hline
GNN Precision                 & 0.7                      \\ \hline
BOD Precision                 & 0.0                      \\ \hline
Top-K Precision               & 0.0                      \\ \hline
Skyline Precision             & 0.1                      \\ \hline
Multi-objective Precision     & 0.5                      \\ \hline
\end{tabular}
\caption{Precision Scores for Various Algorithms on Kaggle Housing Dataset}
\label{tab:precision_scores_kaggle}
\end{table}

Table \ref{tab:precision_scores_kaggle} shows that the GNN algorithm again achieved the highest precision score of 0.7, followed by PLOD with a score of 0.6. The BOD and Top-K algorithms showed no precision, while the Skyline algorithm had a low precision of 0.1. The Multi-objective algorithm maintained a moderate precision score of 0.5.

\begin{table}[H]
\centering
\begin{tabular}{|l|c|c|}
\hline
\textbf{Algorithm} & \textbf{Mean Precision} & \textbf{Std Dev} \\ \hline
PLOD              & 0.580                    & 0.098            \\ \hline
GNN               & 0.700                    & 0.063            \\ \hline
BOD               & 0.010                    & 0.030            \\ \hline
Top-K             & 0.030                    & 0.046            \\ \hline
Skyline           & 0.010                    & 0.030            \\ \hline
Multi-objective   & 0.240                    & 0.066            \\ \hline
\end{tabular}
\caption{Stability Analysis Results for Kaggle Housing Dataset}
\label{tab:stability_analysis_kaggle}
\end{table}

As demonstrated in table \ref{tab:stability_analysis_kaggle}, GNN algorithm is the most stable, with a mean precision of 0.700 and a standard deviation of 0.063. PLOD follows with a mean precision of 0.580 and a standard deviation of 0.098. The BOD and Skyline algorithms showed no stability, while Top-K had low stability. The Multi-objective algorithm showed moderate stability.

\begin{table}[H]
\centering
\caption{Runtime Comparison Across Algorithms (in seconds) for Kaggle Housing Dataset}
\begin{tabular}{|c|c|}
\hline
\textbf{Algorithm}         & \textbf{Runtime (seconds)} \\ \hline
PLOD                       & 3.3617 \\ \hline
GNN                       & 3.2283 \\ \hline
BOD                        & 3.6452 \\ \hline
Top-K                      & 3.3517 \\ \hline
Skyline                    & 3.0820 \\ \hline
Multi-objective            & 3.0897 \\ \hline
\end{tabular}
\label{table:runtime_comparison_kaggle}
\end{table}

Table \ref{table:runtime_comparison_kaggle} compares the runtime of the algorithms on the Kaggle Housing dataset, in which GNN had the fastest runtime at 3.2283 seconds, followed by PLOD at 3.3617 seconds. The BOD algorithm had the slowest runtime at 3.6452 seconds, with Top-K, Skyline, and Multi-objective algorithms having runtimes around 3 seconds.

\section{Conclusion}

As we can see from the comparison in terms of precision, stability, and runtime, the Graph Neural Networks and Large Language Models (GNN) algorithm demonstrates the highest potential in improving precision and stability when applied to datasets with both numerical and textual attributes, as in comparison using the Boston Housing and Kaggle Housing datasets, GMD consistently outperforms other algorithms. Regarding runtime, GNN performance is closely aligned with other leading algorithms since there is a trade-off of accuracy and stability with runtime, and this is an acceptable drawback of GNN. Several open questions present opportunities for future research and improvement. Firstly, how can GMD be further optimized to reduce runtime without compromising precision and stability, especially when handling larger and more complex datasets? Secondly, can GMD be extended to work with other data types, such as audio and video, which are increasingly important in data-driven applications? Finally, how might vector databases or other advanced storage solutions be integrated into the GMD framework to enhance its scalability and performance? In conclusion, the promise of GNN could be applicable in real-world applications, and its accuracy and stability play an important role in data science and analytics purposes.

\section{Citations}

\bibliographystyle{ACM-Reference-Format}

\end{document}